\documentstyle[12pt, aaspp4]{article}

\def\ltsima{$\; \buildrel < \over \sim \;$}
\def\simlt{\lower.5ex\hbox{\ltsima}}
\def\gtsima{$\; \buildrel > \over \sim \;$}
\def\simgt{\lower.5ex\hbox{\gtsima}}

\begin{document}


\title{Effects of Wandering on the Coalescence of Black Hole Binaries
in Galactic Centers}

\author{Pinaki Chatterjee$^1$, Lars Hernquist$^2$ \& Abraham Loeb$^3$}
\affil{Harvard-Smithsonian Center for Astrophysics, 60 Garden Street,
Cambridge, MA 02138}
\footnotetext[1] {pchatterjee@cfa.harvard.edu}
\footnotetext[2] {lars@cfa.harvard.edu} 
\footnotetext[3] {Guggenheim fellow. On sabbatical leave at the
Institute for Advanced Study, Princeton, NJ 08540; loeb@ias.edu.}

\medskip

\begin{abstract}
We examine whether massive binary black holes in spherically symmetric
bulges of galaxies can achieve coalescence through the emission of
gravitational radiation under the action of stellar dynamical
processes alone. In particular, we address the importance of the
Brownian motion of a binary's center-of-mass to its continued
interaction with stellar orbits that allows it to keep hardening:
the restoring force holding the binary at the center
decreases as it depletes the central stellar density, causing the
amplitude of wandering to increase.
We use an analytical model and N-body simulations to calculate the time
required to reach the gravitational-radiation dominated stage.  We
find that a substantial fraction of all massive binaries in galaxies
can coalesce within a Hubble time.
\end{abstract}


\section{Introduction}

Observations indicate that most galaxies harbor supermassive black
holes (BHs) at their centers (e.g., Magorrian et al. 1998). Since
galaxy mergers are inevitable, there arises the question of the dynamical
fate of the binary black holes (BBHs) that result from these mergers.
This problem was first addressed by Begelman, Blandford \& Rees
(1980) (for recent work, see, e.g., Milosavljevi\'c \& Merritt 2001,
2002, Yu 2002).
Dynamical friction causes the black holes in the original
galaxies to sink to the center of the merged galaxy, and then become
bound in a binary system. The semi-major axis of the binary, $a$,
continues to shrink owing to dynamical friction until it reaches the
point at which the binary becomes hard: $a_h \simeq Gm_2/4 \sigma^2$,
where $\sigma$ is the one-dimensional stellar velocity dispersion, $G$
is the gravitational constant, and $m_2$ ($<m_1$) is the mass of the
lighter of the two black holes (following Quinlan 1996). After this
stage, the dominant mechanism by which the binary shrinks and loses
energy is by three-body interactions with stars that pass within a
distance $\sim a$ of the center of mass of the binary.  The binary
transfers energy to the stars by ejecting them at a much higher
velocity. Once it shrinks to the point at which energy loss to
gravitational waves becomes dominant, the binary can quickly coalesce.

Begelman, Blandford \& Rees (1980) conjectured that hardening of a
tight binary by three-body interactions would continue only until it
had interacted with all the stars contained within its loss cone
(Frank \& Rees 1976); if this was not sufficient to drive it to the
gravitational wave emission stage, the binary would stall at a certain
separation, and harden very slowly at the two-body relaxation
timescale (the timescale at which the loss cone can be repopulated),
which turns out for many actual galaxies to be longer than the Hubble
time. The lack of observational evidence for the existence of massive
BBHs has led to suggestions that BBH merging may be facilitated by
other processes, such as the loss of angular momentum to gas (e.g.,
Begelman, Blandford \& Rees 1980, Gould \& Rix 2000, Armitage \&
Natarajan 2002), or triaxiality of the host galaxy (Yu 2002).

It is therefore of interest to inquire whether stellar dynamical
processes alone would be sufficient to drive the BBH to the
gravitational wave emission stage. Quinlan \& Hernquist (1997) pointed
out the importance of the wandering of the binary to its hardening. In
this paper, we investigate the shrinking of a hard black hole binary
in a spherically symmetric galaxy, and ask particularly how Brownian
motion of its center of mass can contribute to its continued hardening
beyond the stage of loss cone evacuation. In \S 2, we set up an
analytic model that appears to account for the results of N-body
simulations described in \S 3. We comment on the impact of BBH
wandering in \S 4, and derive characteristics of the merger in \S 5,
on the assumption that hardening proceeds at the rate deduced in \S 2
and \S 3. We summarize our findings in \S 6.

\section{The Model}

Consider two black holes of masses $m_1$ and $m_2$. Let
$\mu_{12}=m_1m_2/(m_1+m_2)$ be the reduced mass of this BBH
system. Also, let $m_2$ be the lighter of the two BHs, with the mass
ratio being $q = m_2/m_1 \leq 1$.

For illustrative purposes,
we take the host cluster of stars to be described by a Plummer model
of total mass $M$ and length parameter $r_0$.  Thus, the density and
potential profiles are given, respectively, by
\begin{equation} \label{dens}
\rho (r) = \frac{3 M r_0^2}{4 \pi} \frac{1}{(r^2+r_0^2)^{5/2}} ,
\end{equation}
\begin{equation} \label{pot}
\Phi (r) = -\frac{G M}{(r^2+r_0^2)^{1/2}} ,
\end{equation}
where $G$ is the gravitational constant and $r$ is the radial
distance from the center of the stellar system, which is taken
as the origin. In this case, the one-dimensional stellar velocity
distribution in the core is given by $\sigma^2=GM/6r_0$.

The phase space distribution function of the stellar system, $f$,
depends in general both on position $\vec{r}$, and stellar velocity
$\vec{v_a}$, and is defined such that $f({\vec{r}}, {\vec{v_a}})\,
d^3{\vec{r}}\,d^3{\vec{v_a}}$ is the mass in stars in the phase space
volume $d^3{\vec{r}}\, d^3{\vec{v_a}}$.  For the spherically symmetric
Plummer model, we assume
$f$ to be a function of the relative energy per unit mass
${\mathcal E}$ only (and independent of specific angular momentum),
where ${\mathcal E} = - \frac{1}{2} v_a^2 - \Phi (r) = \Psi (r) -
\frac{1}{2} v_a^2$, $\Psi (r) = - \Phi (r)$ being the relative
potential. The distribution function is then (e.g., Binney \& Tremaine
1987)
\begin{equation} \label{dist}
f({\mathcal E})=\frac{96}{7 \sqrt{8} {\pi}^3} \frac{M r_0^2}{(G M)^5} 
{\mathcal E}^{7/2}.
\end{equation}

Consider an encounter between the BBH and a star of mass $m_*$. Let
the impact parameter be $p$ and the initial relative velocity (when
the star is at infinity) be $V_0$. If $r_{min}$ is the minimum
distance reached by the star to the BBH, and $V_{max}$ is the velocity
at that point, then the angular momentum is $l=p V_0=r_{min}
V_{max}$. The reduced mass of the BBH-star system is $\mu \sim m_*$,
assuming the BHs to be much more massive than the stars. Then by
conservation of energy,
\begin{equation} 
E=\frac{1}{2} m_* V_{max}^2- \frac{G (m_1+m_2) m_*}{r_{min}}=
\frac{1}{2} m_*
V_0^2
\end{equation}
Eliminating $V_{max}$ and rearranging,
\begin{equation} \label{p}
p^2=r_{min}^2+\frac{2 G (m_1+m_2)}{V_0^2} r_{min}.
\end{equation}
We set $r_{min}=a$, the semi-major axis of the BBH, implying that an
encounter occurs (i.e., according to our picture, the star interacts
with the BBH and gets thrown out by the slingshot effect, while $a$
shrinks) if $p^2 \le a^2+\frac{2 G (m_1+m_2)}{V_0^2} a$.

The number of stars per unit volume with velocity $v_a$ is $f(v_a)
d^3v_a/m_*$. Hence, the number of encounters per unit time is
\begin{equation} \label{rate}
\frac{1}{t_{en}}=\int \frac{f(v_a)}{m_*} d^3v_a \pi p^2 V_0,
\end{equation}
where $V_0=| \vec{v}-\vec{v}_a |$, $\vec{v}$ being the velocity of the
BBH. In order to take into account the wandering of the BBH's center
of mass, this expression must be averaged over $\vec{v}$ (for a
similar calculation, see Binney \& Tremaine [1987], \S 8.5).  It was
shown in Chatterjee, Hernquist \& Loeb (2002a, b) that the
distribution of each component of $\vec{v}$ is Gaussian, so that the
distribution of the magnitude, $v$, is $W(v)=\frac{2 v^2}{\sqrt{2 \pi}
\sigma_{bh}^3} e^{-v^2/2 \sigma_{bh}^2}$, with
$\sigma_{bh}^2=\frac{2GMm_*}{9r_0 (m_1+m_2)}$.
Thus, averaging over $W$, 
\begin{equation}
\frac{1}{t_{en}}=\frac{2 \pi}{\sqrt{2 \pi} m_* \sigma_{bh}^3} \int v^2
e^{-v^2/2 \sigma^2} \Bigg(a^2+\frac{2G(m_1+m_2)}{V_0^2} a\Bigg) V_0 f(v_a) 
d^3v_a dv,
\end{equation}
or
\begin{equation} \label{ten}
\frac{1}{t_{en}}=\frac{2 \pi}{\sqrt{2 \pi} m_* \sigma^3} \int v^2
e^{-v^2/2 \sigma^2} \Bigg(a^2 |\vec{v}-\vec{v_a}| + \frac{2G(m_1+m_2)}
{|\vec{v}-\vec{v_a}|} a\Bigg) A (\psi-v_a^2/2)^{7/2} d^3v_a dv,
\end{equation}
where $A=\frac{96}{7 \sqrt{8} {\pi}^3} \frac{M r_0^2}{(G M)^5}$. Note
that in the core, $\psi$ obtains an approximately constant value,
$\psi \sim GM/r_0$.

The above integral can be simplified using the approximation
$|\vec{v}-\vec{v_a}| \sim v_a$, since the BBH moves slowly relative to
the stars when its mass is much larger than that of an individual
star.  The integration then gives,
\begin{equation}
\frac{1}{t_{en}}=\frac{32 \pi^2 A \psi^{11/2}}{99 m_*} \Bigg(a+\frac{11}{2}
\frac{G(m_1+m_2)}{\psi}\Bigg)a.
\end{equation}

The BBH's binding energy is $E=\frac{Gm_1m_2}{2a}$. If each star on
average extracts energy $\Delta E$, where $\Delta E=G \mu_{12} m_*/a$,
is roughly the potential energy of the BBH-star system at the point of
closest approach (giving $\Delta E/E=\frac{2 m_*}{m_1+m_2}$), then
\begin{equation}
\frac{\Delta E}{\Delta t}= \frac{Gm_1m_2m_*}{m_1+m_2} \frac{32 \pi^2 A 
\psi^{11/2}}{99 m_*}\Bigg(a + \frac{11}{2} \frac{G(m_1+m_2)}{\psi}\Bigg)
= -\frac{Gm_1 m_2}{2a^2} \frac{da}{dt},
\end{equation}
and simplifying,
\begin{equation} \label{form}
-\frac{1}{a^2\Big(a+\frac{11G(m_1+m_2)}{2\psi}\Big)} \frac{da}{dt}=
\frac{64 \pi^2 A \psi^{11/2}}{99(m_1+m_2)} .
\end{equation}

Thus, there are two limiting regimes for the temporal evolution of $a$:
\begin{itemize}
\item If $a \gg 11G(m_1+m_2) /2\psi$, then $\frac{-1}{a^3} \frac{da}{dt} 
=\frac{C}{m_1+m_2}$,
where $C$ is a constant, and we get $1/a \propto t^{1/2}$. 

\item If $a \ll 11G(m_1+m_2) /2\psi $, then $\frac{d}{dt}(\frac{1}{a})=$
a constant, and we get $1/a \propto t$.

\end{itemize}
For intermediate values, we get a power law, $1/a \propto t^d$,
with $0.5 \leq d \leq 1.0$.

Now, $11G(m_1+m_2)/2\psi = \frac{11}{2} \frac{m_1+m_2}{M} r_0$. 
The BBH becomes hard at $a_h \sim \frac{3}{2} \frac{m_2}{M} r_0$, where
$m_2$ is the lighter of the two BHs. 
Thus, after the BBH becomes hard, we
always tend to the second case above, $1/a \propto t$.

In our numerical experiments (to be described in the next section), we
have found temporal power-law indices for $1/a$ which are always
bounded between $\sim 0.5$ (for heavy BBHs) and 1 (for light BBHs).
The above simplified theory predicts a power-law index of 1 for hard
binaries. However, our experiments showed that heavy BBHs (which
contain more than a few percent of the mass of the Plummer galaxy)
cause the central density of the galaxy to drop quickly, and so it is
not appropriate to assume, as we did above, that the stellar
distribution function remains unaffected by the BBH. In particular,
the left-hand-side of equation (\ref{form}) needs to be
$\frac{-1}{a^2(a+c_1)} \frac{da}{dt}$, where $c_1$ is now some
undetermined term. If $c_1$ is not small compared to $a$, then it is
possible to get $1/a \propto t^d$, with $d < 1$. This description is
suggestive for why we obtained such limiting behaviors, and it is due
to the form of the expression (\ref{p}) for the impact parameter being
a sum of two terms.

For low mass BBHs, the distribution function is not much affected by
the hardening of the binary, and so we may proceed to derive the temporal
slope of the $1/a$ line,
\begin{equation} \label{slope}
s=\frac{d}{dt} \Bigg(\frac{1}{a}\Bigg) =\frac{256 \sqrt{2}}{21 \pi}
\Bigg(\frac{GM}{r_0^5}\Bigg)^{1/2}.
\end{equation}
Note that this result is independent of the masses of the two BHs.

We may compare the above value with the hardening rate presented in 
other work in a slightly different form: $\frac{d}{dt} 
\big(\frac{1}{a}\big) =\frac{G \rho_0 H}{\sigma}$, where $H$ is a 
dimensionless hardening rate, $\rho_0$ is the central stellar density, 
and $\sigma$ is the one-dimensional stellar velocity dispersion in the 
core. Comparing with the above equation, we obtain $H=9.4$. This is very 
similar to the hardening rate observed in simulations by Milosavljevi\'c 
\& Merritt (2001). It also compares well with the value $H \sim 15$ 
(independent of BH mass ratio and orbital eccentricity), which 
is the hardening rate derived from three-body scattering experiments in 
the limit of a very hard binary (Quinlan 1996, Hills 1983).

In our simulations, we made the choice $r_0=3 \pi/16$ and $GM=1$, for
which one gets $s=\frac{d}{dt} (\frac{1}{a}) =20.6$. Note, however,
the approximations that have been made in the derivation. We have
assumed that the distribution function of the stars remains unchanged,
which is clearly false: the central density of the stellar system
falls, as shown in the next section. For low mass BBHs (say $m <
0.01$), however, the decline is not as much as for heavy BBHs, and we
expect the inverse semi-major axis of the BBH to increase at a
constant rate, with $s$ given roughly by the expression above.

We have assumed $\Delta E/E=2m_*/(m_1+m_2)$, a form
which is in reasonably good agreement with results of scattering
experiments at several mass ratios (see Roos 1988, Hills 1983).
It is, of course, an expression for the
average energy extracted by a passing star, and need not accurately
reflect the actual process of energy extraction for all impact
parameters. Given that the
precise form of this expression suitable for use in the massive BBH
case is uncertain (for example, Hills \& Fullerton [1980] find that
for equal-mass binaries, the right-hand side of the above expression
may be multiplied by 0.7), it is perhaps 
optimistic to expect this derivation to give the correct slope $s$. 

\section{Results from Numerical Experiments}

The numerical simulations were carried out using the SCFBDY program,
which is described in detail in Quinlan \& Hernquist (1997) and
Chatterjee, Hernquist \& Loeb (2002a). In our code, each BH interacts
with the other BH and the stars through the $1/r^2$ force, but stars
interact with other stars through a mean field expansion of the
gravitational potential that is updated self-consistently with time
(note that the simulation thus does not treat star-star
relaxation accurately [see Hernquist \& Barnes 1990]; this is 
justified since the two-body relaxation timescale
in real galaxies is much longer than the Hubble time). The particles
are moved with individual step-sizes using the fourth-order
integrators of Aarseth (1994): NBODY1 for the stars and NBODY1 or
NBODY2 for the BHs.

The mean field expansion of the gravitational potential is carried out
in a set of basis functions that includes both spherical and
non-spherical terms (Hernquist \& Ostriker 1992). However, since the
galaxy remained nearly spherical during the integration, it made not
much difference to the hardening of the BBH whether a few or zero 
non-spherical terms were included in the potential expansion. 

We verified that numerical relaxation was not an important factor in
our simulations by forcing the stars to interact only with a fixed
background potential, and not with each other, and finding our results
to be similar to those obtained using the full SCFBDY code.  The 
insignificance of numerical relaxation in our simulations is further 
supported by experiments in which we forced the center of mass of the
BBH to remain fixed, as described below.

Our results are presented in a system of units in which $G=M=1$, and
$r_0=3 \pi/16$, so that the energy of the initial galaxy without the
BHs is -1/4 (Heggie \& Mathieu 1986). The stars (of equal mass
$m_*=1/N$, where $N$ is the number of stars) are initially distributed
according to the Plummer distribution function, and the black holes
are started out on nearly circular orbits at $\vec{r}$ and $-\vec{r}$,
with $r=0.3$. (We tried a range of initial conditions for the BHs,
including elliptical orbits; it made no difference to the hardening of
the binary.)  We have done a range of experiments, with binary mass
ranging from $m_1=m_2=0.00125$ to $m_1=m_2=0.01$, and with particle
numbers of up to $N=400,000$.

Figure 1 shows the results of such an integration with
$m_1=m_2=0.00125$, and 400,000 stars. Shown are the separation between
the BHs, $R$, the inverse semi-major axis of the BBH orbit, $1/a$, the
eccentricity of the orbit, $e$, and the central density, $\rho$, of the
stellar system averaged over a sphere containing a mass in stars
somewhat larger than the mass of the BBH. The BHs become bound at
$t=77.3$. The BBH becomes hard once $a$ shrinks to the value $a_h=Gm_2/4
\sigma^2=1.104 \times 10^{-3}$, or when $1/a_h=905$. As can be seen,
the rate of change of $1/a$ is nearly constant after the binary has
become hard; namely, $1/a \propto t^d$ with $d \simeq 1$.

Figure 2 shows the results of a similar integration with
$m_1=m_2=0.02$, and 200,000 stars. The BHs become bound at $t=4.4$,
and the BBH becomes hard when $a=a_h=0.018$, or when
$1/a=1/a_h=56.6$. The rate of hardening is clearly not constant after
this stage. Indeed, this is at the opposite extreme from the previous case, 
with the hardening being well fitted by $1/a \propto t^d$ with $d \sim
0.5$. 

For intermediate values of $m_1$ and $m_2$, the hardening occurs at
the rate $1/a \propto t^d$ with $0.5 \simlt d \simlt 1.0$. We have not
found values of $d$ significantly outside this range. This is
consistent with the explanation given in \S 2. Notice that for
$m_1=m_2=0.00125$, the density at the position of the BBH center of
mass drops gently over the course of its hardening. This is expected
since the BBH expels stars from the center as it hardens. The
magnitude of the drop is small, however, and thus to a first
approximation, the stellar distribution function remains unaffected.
On the other hand, the heavy BBH changes the distribution function to
a greater extent, as is clear from the bottom right panel of Figure
2. Thus, in the former case, the theory developed in the previous
section holds, and the binary hardens at a constant rate; in the
latter case, we get a qualitatively different behavior, for the reasons
outlined in \S 2.

The rate of hardening over time is close to constant in three cases
for which we ran simulations:~$m_1=m_2=0.00125$, $m_1=m_2=0.0025$, 
and ~$m_1=m_2=0.005$. Like Quinlan \& Hernquist (1997), we find that the
slope of the $1/a$ line, given by $s=\frac{d}{dt}(1/a)$, varies with
the number of stars. As the number of stars is increased, $s$ falls
systematically, until there are roughly 200,000 -- 400,000 stars, when
it stabilizes to a particular value, $s_0$. This decrease in the slope
as the number of stars increases has also been observed by Makino
(1997) and Hemsendorf, Sigurdsson \& Spurzem (2002), though not by
Milosavljevi\'c \& Merritt (2001). Quinlan \& Hernquist (1997)
explained it as resulting from two competing factors governing the
wandering of the BBH's center of mass: on the one hand, as the number
of particles, $N$, rises, the binary wanders less from the center, and
thus it reduces the density at the center more, and the hardening rate
decreases; on the other hand, the restoring force of the stellar
potential goes down as the central stellar density drops, causing an
increase in the BBH's wandering. At a particular value of $N$, there
is a balance between these two factors, and the hardening rate stops
changing for higher values of $N$.

Quite apart from the variation of $s$ with $N$, we found that the
limiting value of $s$ changes also with the mass of the BBH.  We found
that for $m_1=m_2=0.005$, $s$ stabilizes to a value of $s_0 \sim 8$;
for $m_1=m_2=0.0025$, $s_0 \sim 11$; and for $m_1=m_2=0.00125$, $s_0
\sim 15$. Recall that $s=20.6$ according to the simplified theory of
\S 2. That value is expected to be obtained in the limit that the
stellar distribution function remains unchanged by the influence of
the binary. Our numerical experiments indeed reveal that this becomes
a better approximation as the binary mass is reduced.  However, the
question remains whether this value is actually reached for still less
massive BBHs.

The orbital eccentricity of the BBH in its hard state varies according 
to the mass of the binary. For massive binaries, such as for the case
$m_1=m_2=0.02$ (Figure 2), the BBH orbits are circular to a good
approximation. For smaller masses of the BBH, the eccentricity is
larger: e.g., for $m_1=m_2=0.00125$ (Figure 1), the eccentricity in
the hard state is roughly 0.3, and trending very slightly
upward. These values are consistent with Quinlan \& Hernquist (1997)
and Milosavljevi\'c \& Merritt (2001), and with the results of 
scattering experiments in the restricted three-body approximation 
(Quinlan 1996), according to which the BBH eccentricity hardly grows if 
the initial eccentricity is moderate ($e_0 \le 0.3$), and does not grow 
by much if it is larger, before the gravitational radiation stage sets 
in. However, Hemsendorf, Sigurdsson \& Spurzem (2002) and Aarseth (2002) 
have recently found that the eccentricity may grow to values 
close to 1, resulting in rapid coalescence by the emission of 
gravitational radiation (see \S 5). The reason for this discrepancy is 
not clear, and may have to do with the initial conditions chosen for the 
simulations.

It is of interest to know the extent of wandering of the
BBH. According to the formulae developed in Chatterjee, Hernquist \&
Loeb (2002 a, b), the mean squared position and velocity of the center
of mass of the BBH in the hard stage should, for the case of the
Plummer potential, be given, respectively, by
\begin{equation} \label{wandr}
r^2_{cm}=\frac{2 r_0^2m_*}{3 (m_1+m_2)},
\end{equation}
\begin{equation} \label{wandv}
v^2_{cm}=\frac{2 GMm_*}{3 r_0 (m_1+m_2)}.
\end{equation} 
This assumes that the stellar distribution function is unchanged,
which is not the case for a massive BBH; in fact, the
wandering should be larger than indicated by these values, since the
restoring force on the BBH drops as the central stellar density is
diluted by the BBH.  However, the above expressions provide a
reasonable description for the case of light BBHs. Indeed, we find
from simulations that for such cases, the wandering is consistently
larger than the values provided by the above formulae, but not by more
than a factor of $\sim 2$ (see Table 1). The reasons for the enhanced
wandering could still be the somewhat reduced stellar density at the
center of the cluster, as well as the enhanced random motion
of a binary compared with that of a point mass, as discussed by
Merritt (2001).  Thus, the binary is able to interact with a larger
region of stellar phase space than calculated according to the simple
model of \S 2.

\section{The Impact of Wandering on the Hardening of the BBH}

Quinlan \& Hernquist (1997) reported an experiment in which the center
of mass of the binary was held fixed at the origin by the application
of an artificial constraint force: they found that hardening halted
soon after the binary became hard, following the evacuation of the
loss cone. We have confirmed this result in our own experiments (see
Figure 2 for an example): the BBH becomes hard when its semimajor axis
reaches $a_h$, and continues hardening by three-body interactions with
the stars until $a=a_{lc}~(a_{lc}<a_h)$, when the loss cone is empty. These
experiments show the importance of wandering in keeping the BBH 
hardening. (Incidentally, the fact that hardening stops in this case
is another indication that numerical relaxation is not causing
spurious refilling of the loss cone in our simulations.)

According to traditional theory (e.g., Begelman, Blandford and Rees
[1980]), hardening of the binary should stop after the BBH has
interacted with all the stars in its loss cone, as long as its
wandering is unimportant. However, there is no sign in our experiments
that a ``hole'' develops in the density distribution around the binary
as it hardens -- instead, the central density drops slowly as the
binary hardens (see Figure 1). On the contrary, as explained in the
previous section, the wandering of the BBH increases as the central
density falls, allowing it to interact with new regions of stellar
phase space, and to keep hardening at a rate consistent with $1/a
\propto t^d$. The value of $d\approx 1$ is realized in the limit where
the BBH makes up a small fraction of the host stellar system mass, as is
typically the case for galactic bulges.

What if the mean wandering radius of the BBH were smaller than the
semi-major axis at which the binary becomes hard ($a_h$)? Then the
binary would wander ``within itself'', so to speak; in this case one
may wonder whether it would not encounter fresh regions of phase space,
and perhaps hardening would stop (the bottleneck, if there is one,
would occur at the stage at which the binary becomes hard, because
wandering only becomes more important as the semi-major axis of the
binary shrinks). Is there a sign of this happening in the simulations?

Unfortunately, to do this experiment in the regime in which we are
most interested (i.e., for BBHs of low mass, so that the hardening
rate is constant over time), one would require the particle number to
be prohibitively high. However, it is possible to perform this
experiment for more massive BBHs. For example, Table 2 lists the
experimental root-mean-square value of the binary's Brownian
motion,~$r_{cm}$, for a number of experiments, with the corresponding
value of $a_h$; in each of the cases, $r_{cm} < a_h$. Hardening does
not stop at or near this point, but continues indefinitely beyond this
value at the rate $1/a \propto t^d$, with $0.5 \simlt d \simlt 1.0$.
Figure 2 shows the hardening of a BBH with
$m_1=m_2=0.02$, with 200,000 particles.  The rms value of the binary's
Brownian motion is 0.013; the binary becomes hard when $a=a_h=3
m_2a/2M=0.018$, or when $1/a_h=56$.

In each of these cases, however, we find that $r_{cm} > a_{lc}$;
moreover, the value of $r_{cm}$ is very different from the value
predicted by equation (\ref{wandr}). This indicates that when the
binary is allowed to wander, hardening no longer stalls at the
semimajor axis $a_{lc}$; rather, the amplitude of its wandering
increases as the central density of the star cluster declines (see
Figure 3 for an explicit example of this), resulting in a lower
restoring force holding the binary at the center. This, in turn,
allows the binary to keep hardening by interacting with new stars that
it would not have encountered if it did not wander. Thus, the ``loss
cone'' is repopulated at a rate higher than the two-body relaxation
rate by virtue of the increased wandering of the BBH's center of
mass.

\section{Implications for the Coalescence of Massive Black Hole 
Binary Systems} 

We have evidence, therefore, that the BBH can continue to harden
beyond the point at which the loss cone would be evacuated were the
binary's center of mass to be fixed at the origin, and that, moreover,
it does so at roughly the rate calculated in \S 2. If we assume that
hardening can continue until the point at which the emission of
gravitational radiation becomes the dominant mode of energy
loss for the binary, we can calculate when that should occur.

From equation (\ref{slope}), the hardening timescale of the BBH is
\begin{equation}
t_h(a)=\Big|\frac{a}{\dot{a}}\Big|=
\frac{21 \pi}{256 \sqrt{2}}\Bigg(\frac{r_0^5}{GM}\Bigg)^{1/2}\frac{1}{a}~~~.
\end{equation}

The hardening timescale for gravitational radiation is (Peters 1964)
\begin{equation}
t_h(a)=\Big|\frac{a}{\dot{a}}\Big|=\frac{5}{64} \frac{c^5a^4}{G^3 \mu_{12}
(m_1+m_2)^2} \frac{(1-e^2)^{7/2}}{1+73e^2/24+37e^4/96},
\end{equation}
where $c$ is the velocity of light, and $e$ the eccentricity of the
BBH's orbit. The dependence of the gravitational radiation timescale
on eccentricity is weak unless $e$ is close to 1. As noted in \S 3,
$e$ is very small for heavy BBHs, but rises to larger values ($\sim
0.2$) for lighter BBHs. In the absence of a precise characterization
of the variation of $e$ with BBH mass, we take $e=0$ in the above
expression; thus, our result for the time of onset of gravitational
wave emission and coalescence should be regarded as an upper limit,
as a higher eccentricity would allow this stage to be reached more
quickly.

Equating the two expressions, we have that the semi-major axis at which
gravitational radiation losses start to dominate, $a_{gr}$, is given
by
\begin{equation}
a_{gr}^5=\frac{21 \pi}{20 \sqrt{2}} \frac{G^{5/2}m_1^3 q(1+q)
r_0^{5/2}}{c^5 M^{1/2}},
\end{equation}
where $q$ is the mass ratio $q=m_2/m_1 \le 1$.

From equation (\ref{slope}), assuming that $a_{gr} \ll a_h$, we can
derive the time between the binary's becoming hard at semi-major axis
$a_h$, and the time at which $a=a_{gr}$,
\begin{equation}
t_{gr}=\frac{21 \pi}{256 \sqrt{2}} \frac{r_0^{5/2}}{G^{1/2} M^{1/2}}
\frac{1}{a_{gr}},
\end{equation}
or
\begin{equation}
t_{gr}=\frac{21 \pi}{256 \sqrt{2}} \Bigg(\frac{20 \sqrt{2}}{21 \pi}
\Bigg)^{1/5}
\frac{c r_0^2}{G M^{2/5} m_1^{3/5}[q(1+q)]^{1/5}}.
\end{equation}
The dependences on $M, m_1$ and $q$ are rather weak.

We can also calculate the total number of stars,~$N_{int}$, the BBH must 
interact with from the time it becomes hard to the time $t_{gr}$. We have
$\Delta E = G \mu_{12}m_*/a=\frac{G m_*}{a} \frac{m_1
q}{1+q}$. Therefore $dE=\frac{G m_*}{a} \frac{m_1
q}{1+q} dN$, or $\frac{dE}{dN}=\frac{G m_*}{a} \frac{m_1
q}{1+q}$. But $E=Gm_1 m_2/2a$, and so $\frac{dE}{dN}=\frac{-Gm_1^2q}{2
a^2} \frac{da}{dN}$. Equating the above, we get
$\frac{dN}{da}=-\frac{m_1}{m_*} \frac{1+q}{2a}$. Integrating, the
total number of stars the BBH interacts with is
\begin{equation}
N_{int}=\frac{m_1 (1+q)}{2 m_*} \textrm{ln}\Big(\frac{a_h}{a_{gr}}\Big),
\end{equation}
where $a_h=3m_2r_0/2M=3m_1qr_0/2M$.
The total mass with which the BBH interacts is then simply 
\begin{equation}
M_{int}=\frac{(m_1+m_2)}{2} \textrm{ln}\Big(\frac{a_h}{a_{gr}}\Big).
\end{equation}
As will be seen from Figure 4c, the total mass in stars the BBH must
interact with is roughly of the order of itself. Therefore, according
to this theory, the BBH should not change the stellar distribution
function much, as long as the mass of the BBH is not significant in
comparison with the total mass in stars.  This result is consistent
with our simulations for small mass BBHs.

Some of the above relations are depicted for physically reasonable
stellar systems and BBHs in Figure 4. Instead of presenting results in
terms of $M$ and $r_0$, we use more physically and observationally
relevant variables, such as the one-dimensional stellar
velocity dispersion, $\sigma$, and the central density of the stellar
system, $\rho_0$. The relations amongst these quantities for the
Plummer model are as follows:
\begin{displaymath}
\rho_0=\frac{3M}{4 \pi r_0^3},
\end{displaymath}
\begin{displaymath}
\sigma^2=\frac{GM}{6r_0}.
\end{displaymath}
We then obtain
\begin{equation}
a_{gr}^5=\frac{63 \sqrt{3}}{80} \frac{G^2m_1^3 q(1+q) \sigma}{c^5 \rho_0},
\end{equation}
and
\begin{equation}
t_{gr}=\frac{63 \sqrt{3}}{1024} \Bigg(\frac{80}{63 \sqrt{3}}\Bigg)^{1/5}
\frac{c \sigma^{4/5}}{G^{7/5} m_1^{3/5}[q(1+q)]^{1/5} \rho_0^{4/5}}.
\end{equation}

Figures 4a and 4b show that for physically reasonable masses of the
BBH and characteristics of the host galaxy, the semi-major axis can
shrink to the value $a_{gr}$ within a Hubble time, after which
coalescence occurs quickly by the emission of gravitational waves. The
actual values of $a_{gr}$ are very small (Figure 4d).

The above calculations assume that the binary hardens according to
$1/a \propto t$, which is the behavior we see in the simulations when
the mass of the BBH is less than about 1\% of the total mass in
stars. Is this a reliable guide to what happens in actual galaxies?

The ratio of the mass of the BH to the mass of the galaxy is observed
to be quite low (on the order of $10^{-3}$: see Merritt \& Ferrarese
2001, Wandel 1999, Magorrian et al. 1998), and therefore the linear
hardening behavior is most likely to apply in the cases of practical
interest.

However, it is interesting to ask what would happen if the BBH
hardened more slowly than at a constant rate. As an example, consider
the case shown in Figure 2, with $m_1=m_2=0.02$, and $N=200,000$
stars. The hardening behavior is well fit by the expression $1/a =21.7
\times t^{0.5}$. For purposes of illustration, let us use $a_{gr_{1}}$
to denote the semimajor axis at which gravitational wave radiation
becomes dominant if we assume that the BBH hardens at a constant rate,
$1/a \propto 20.6 \times t$, and $a_{gr_{2}}$ the semimajor axis at
which gravitational wave radiation becomes dominant if we assume that
the BBH hardens at the above non-linear rate, $1/a \propto 21.7 \times
t^{0.5}$, as is actually observed in the simulations. If we scale our
galaxy such that $r_0=3 \pi/16$ corresponds to 100 pc, and $M=1$
corresponds to a total mass in stars in the bulge of $10^9 M_{\odot}$,
then we obtain $a_{gr_{1}}/a_{gr_{2}} \simeq 0.3$. Thus, the
gravitational radiation stage sets in at a larger semimajor axis if
hardening occurs more slowly than linearly.  This arises from the
different behaviors of the hardening timescales: when $1/a \propto t$,
$t_h = |a/\dot{a}| \propto 1/a$, whereas when $1/a \propto t^{0.5}$,
$t_h \propto 1/a^2$. However, the times taken to reach $a_{gr_{1}}$
and $a_{gr_{2}}$ are approximately 1 Gyr and 70 Gyr respectively,
since in the latter case the BBH hardens only as $t^{0.5}$ rather than
$\propto t$, as in the former case. Therefore, despite the fact that the
gravitational radiation stage is reached at a larger semimajor axis if
hardening occurs slower than linearly, the time taken to reach this
stage is longer in the former case. Thus, our results in this section
hold for light BHs which, as mentioned before, is the case with most
galaxies.

\section{Summary}

In summary, our primary results are:

\begin{itemize}

\item We have developed a consistent description for the behavior of
the semi-major axis of a hard BBH. As long as the BBH mass is small
compared to the total mass in stars: $1/a \propto t^d$, with $d\approx
1$.

\item Our simplified theory gives the value of the proportionality
constant of the above linear relation, $s=20.6$, independent of the
mass of the BBH or the number of stars. We find, however, from
numerical experiments that $s$ depends on the number of stars -- as
the number of stars increases, $s$ drops, but eventually converges to
a particular value for $N\ga 200,000$ stars. A possible explanation
is that this reflects an equilibrium between two competing processes:
as the number of stars increases, the wandering of the binary becomes
smaller, causing a larger drop in central density; this, however,
increases the wandering of the BBH because of the reduction of the
central restoring force on it.

\item We also find that $s$ depends on the mass of the BBH: in our
simulations for equal-mass BHs, the convergent value of $s$ increases
from $\sim 8$ to $\sim 15$ as the mass of each BH drops from 0.005 to
0.00125. We interpret this as meaning that the theoretical value 20.6
is reached when no change in the stellar distribution function is
caused by the BBH -- this will be true only when the mass of the BBH
is very small compared to the total mass in stars. If we could run
simulations with even lower mass BBHs, we may approach 20.6. However,
it should be remembered that we have assumed a somewhat
uncertain form for the energy removed from the binary by each stellar
interaction. We also note that for calculating the time at which
gravitational radiation starts to dominate the energy loss, we do not
need to know the precise value of $s$, since the fifth root of this
number needs to be taken, as shown in \S 5.  There are two
difficulties with running simulations with very light BBHs: first, the
black holes need to be much more massive than the stars; second, the
dynamical friction timescale for light black holes, and consequently the time
for them to become hard, is very long.

\item For light BBHs, strong disruption of the stellar density does
not occur: the central density of the cluster does decline with time
but gently. This is where we expect our theory to be applicable. There
is no sign that a ``hole'' is forming at the center, and thus no sign
that hardening would stop after the loss cone is evacuated. Wandering
allows the BBH to interact with regions of stellar phase space it
would not otherwise encounter. According to our experiments, if we
keep the BBH center of mass artificially clamped to the center,
hardening quickly stops after the BBH has interacted with the stars
that could pass within roughly a distance $a$ from it.

\item For heavy BBHs, there is stronger disruption of the stellar density
    and distribution function, and strong deviations from the linear
    behavior of the inverse semi-major axis:  $1/a \propto t^d$, with
    $0.5\la d\la 1$. This is the
    regime in which our theory is not applicable, though it
    gives a qualitative explanation for why $d$ should be bracketed by
    0.5 and 1.

\item We have shown that the hardening of the BBH does not stop even
when its typical wandering radius is smaller than the value of the
semi-major axis at which it becomes hard, at least in the case when
the BBH is fairly massive; however, the wandering radius is always 
larger than the semi-major axis at which hardening would halt because 
of loss cone depletion if the binary did not wander. The reason involves 
the increased wandering of the binary's center of mass as the central 
density of the stellar system falls, resulting in a reduction of the 
restoring force keeping the binary at the center.  To verify this in the 
case of light binaries would require simulations with many millions of 
particles.

\item We have calculated an upper limit on the time required for the
BBHs to reach the stage at which energy losses are dominated by
gravitational radiation emission. For reasonable characteristics of
the stellar system, the gravitational radiation stage is reached
within a Hubble time. In order to reach this stage, the BBH must
interact with a total mass in stars that is of order its own mass.
 
\end{itemize}

Our model uses the spherically symmetric Plummer density profile, 
which has a flat density structure at the center. 
However, observations have shown that most
elliptical galaxies are characterized by weak or strong cusps (e.g.,
Faber et al. 1997, Byun et al. 1996, Lauer et al. 1995). Is it then
reasonable to use the Plummer model in calculations of BBH coalescence?

The Plummer model is characterized by a simple distribution function,
which makes it possible to obtain analytic results for the phenomena
discussed in this paper. Here, we are mainly interested in
understanding the effect of the binary's Brownian motion on its
hardening, rather than in calculating precisely the time taken for a
binary in a real galaxy to coalesce.  Despite the fact that our model
for galactic bulges is unrealistic, we believe that our results
capture generic dynamical processes that take place in more
complicated situations for a number of reasons.
First, recent simulations have shown that the
merger of two galaxies with steep density cusps produce a galaxy with
a shallow density cusp (Milosavljevi\'c \& Merritt 2001), owing to the
transfer of energy from the binary to the stars; moreover, the
destruction of the steep cusp takes place very rapidly (on a timescale
comparable to the local dynamical time), before the binary becomes
significantly hard (similar results have been obtained by Nakano and
Makino 1999 a, b). Our results apply to the stage when the binary has
become hard. Secondly, the weak cusp of the merged galaxy is carried
by the BBH as it wanders around; whereas the stellar density and
velocity dispersion in our calculations refer to their values in the
core of the galaxy but outside the Keplerian rise around the black
holes, since most of the encounters that remove energy from the BBH
are with stars from outside the cusp. Thirdly, the presence of the
density cusp would only increase the supply of stars to the binary,
resulting in more efficient hardening; if a binary coalesces within a
Hubble time in our model, it would likely coalesce within a shorter
time in the presence of a cusp.

\bigskip
\bigskip

We thank G. Quinlan for providing the simulation code. For useful
comments, we thank G. Rybicki and the participants of the Massive
Black Hole Coalescence Focus Session at Pennsylvania State University,
November 2002. AL acknowledges support from the Institute for Advanced
Study at Princeton, the John Simon Guggenheim Memorial Fellowship, and
NSF grants AST-0071019, AST-0204514.

\begin{table}
\begin{center}
\begin{tabular}{|c|c|c|c|c|c|}                              \hline
$m_1=m_2$ & $N$ & $r_{cm}$~(theory) & $r_{cm}$~(expt.) & $a_h$  \\ \hline
0.00125 & 400000 & 0.0152 & 0.0251 & 0.011 \\ \hline
0.00125 & 200000 & 0.0215 & 0.0386 & 0.011 \\ \hline
0.0025 & 200000 & 0.0152 & 0.0298 & 0.022 \\ \hline
0.0025 & 100000 & 0.0215 & 0.0321 & 0.022 \\ \hline
0.005 & 200000 & 0.0108 & 0.0185 & 0.044 \\ \hline
0.005 & 100000 & 0.0152 & 0.0273 & 0.044 \\ \hline
\end{tabular}
\end{center}
\begin{center}
\caption{Values of the wandering radius, $r_{cm}$, of the binary's
center of mass according to our simplified theory (equation
[\ref{wandr}]) and numerical experiments, and the corresponding values
of $a_h$.}
\end{center}
\label{table:1}
\end{table}

\begin{table}
\begin{center}
\begin{tabular}{|c|c|c|c|c|c|}                              \hline
$m_1=m_2$ & $N$ & $r_{cm}$~(theory) & $r_{cm}$~(expt.) & $a_h$ & 
$a_{lc}$ \\ \hline
0.02 & 200000 & 0.0054 & 0.0129 & 0.0177 & 0.008 \\ \hline 
0.04 & 100000 & 0.0054 & 0.0202 & 0.0353 & 0.011 \\ \hline
0.05 & 50000 & 0.0068 & 0.0142 & 0.0442 & 0.013 \\ \hline
0.05 & 100000 & 0.0048 & 0.0321 & 0.0442 & 0.015 \\ \hline
0.10 & 50000 & 0.0048 & 0.0376 & 0.0884 & 0.021 \\ \hline
0.10 & 100000 & 0.0034 & 0.0444 & 0.0884 & 0.021 \\ \hline
\end{tabular}
\end{center}
\begin{center}
\caption{Values of the wandering radius, $r_{cm}$, of the binary
according to our simplified theory (equation [\ref{wandr}]) and
numerical experiments, and the corresponding values of $a_h$ and
$a_{lc}$.}
\end{center}
\label{table:2}
\end{table}

\begin{figure}[t]
\centerline{\epsfysize=6.0in\epsffile{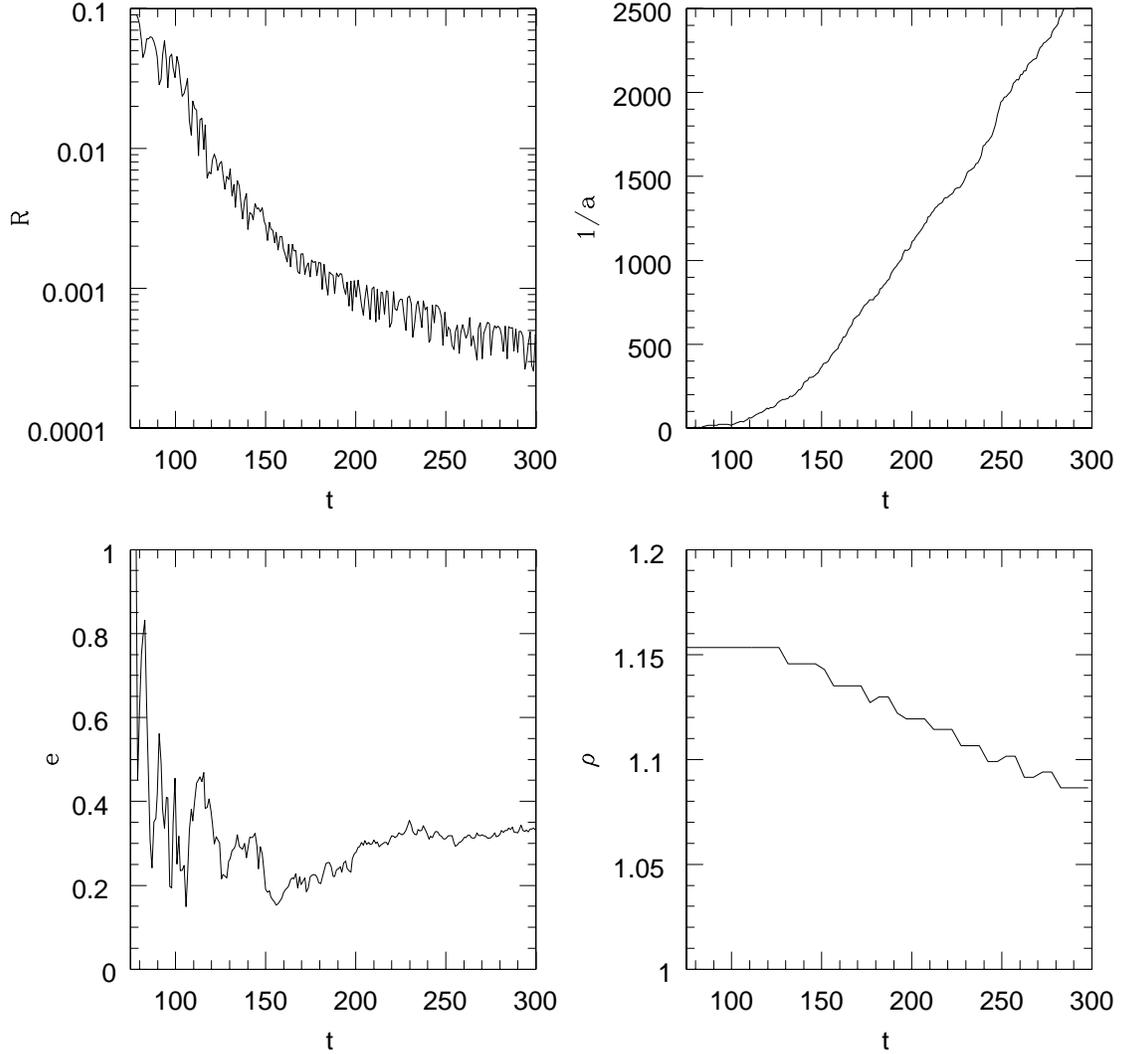}}
\caption{ Results from a simulation of a binary black hole system with
$m_1=m_2=0.00125$ and 400,000 stars. Shown are the black hole
separation, $R$, the reciprocal of the semi-major axis of the binary's
orbit, $a$, its eccentricity, $e$, and the density at the center of
the potential, $\rho$. The binary becomes hard at $1/a \simeq 1/a_h =
905$; it subsequently hardens at approximately a constant rate,
$s=\frac{d}{dt}\big(\frac{1}{a}\big) \simeq 15.5$.  }
\label{fig:1}
\end{figure}

\begin{figure}[t]
\centerline{\epsfysize=6.0in\epsffile{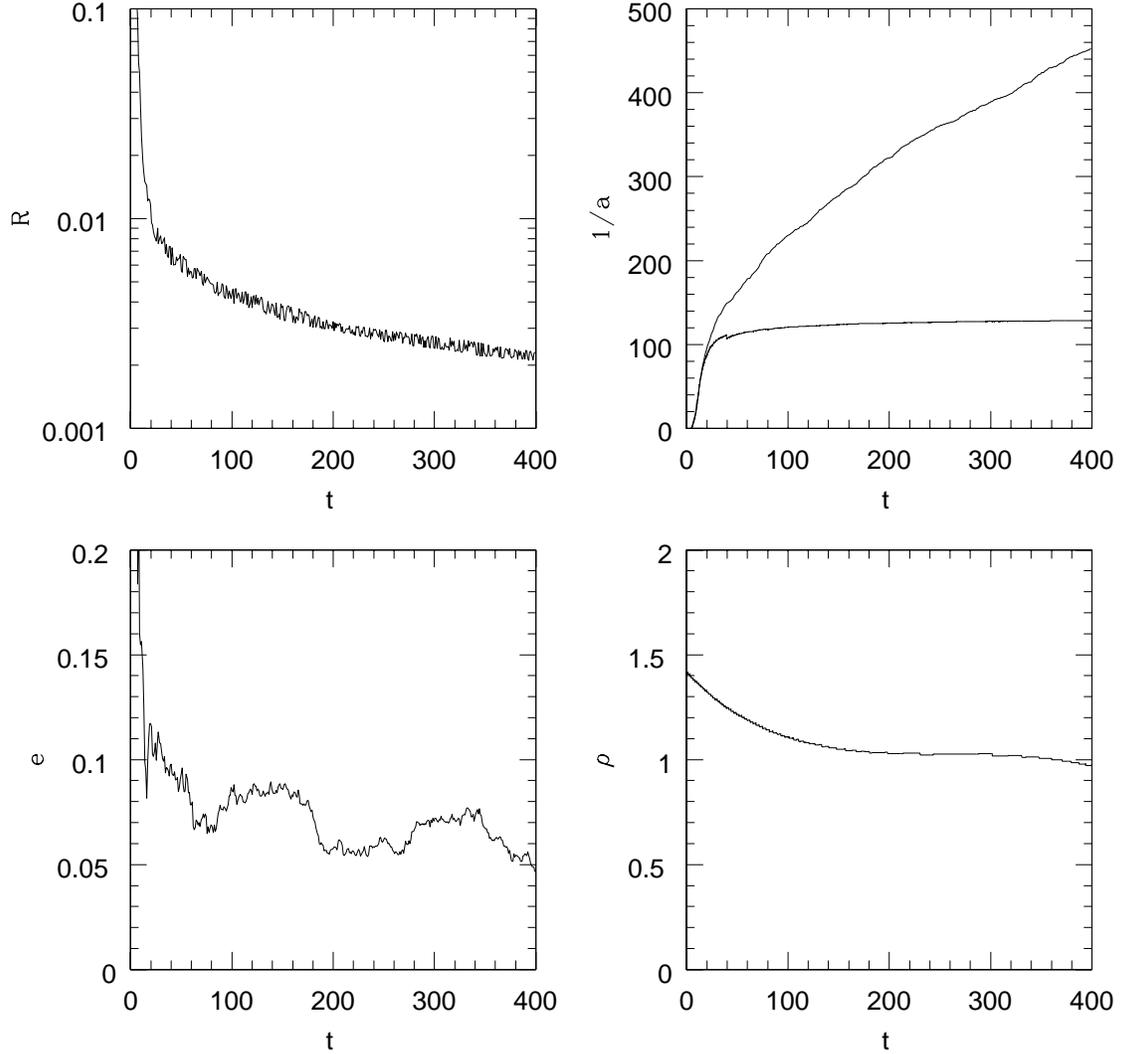}}
\caption{ The same as Figure 1, except that $m_1=m_2=0.02$ and
$N=200,000$.  The binary becomes hard at $1/a \simeq 1/a_h = 57$; it
subsequently hardens at a rate given approximately by $1/a \propto
t^d$, with $d \sim 0.5$. If the center of mass of the binary is held
fixed at the origin, hardening stops as $1/a \simeq 1/a_{lc} = 130$,
as indicated by the lower curve of the upper right panel.  }
\label{fig:2}
\end{figure}

\begin{figure}[t]
\centerline{\epsfysize=6.0in\epsffile{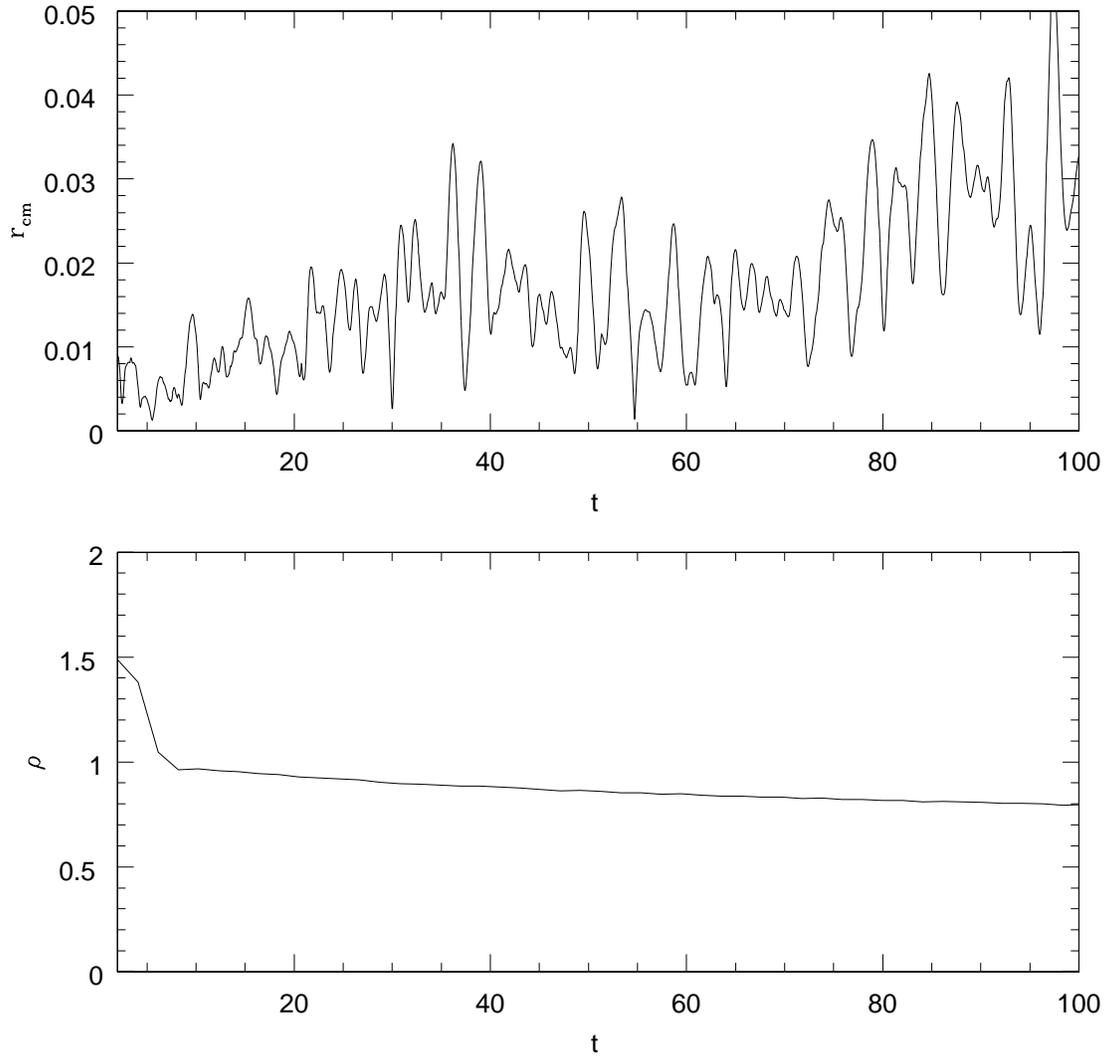}}
\caption{
The radial distance of the binary's center of mass from the origin,
and the density at the center of the potential, for a simulation with 
$m_1=m_2=0.04$ and $N=100,000$. The wandering of the BBH's center of
mass increases as the central density drops.
}
\label{fig:3}
\end{figure}

\begin{figure}[t]
\centerline{\epsfysize=6.0in\epsffile{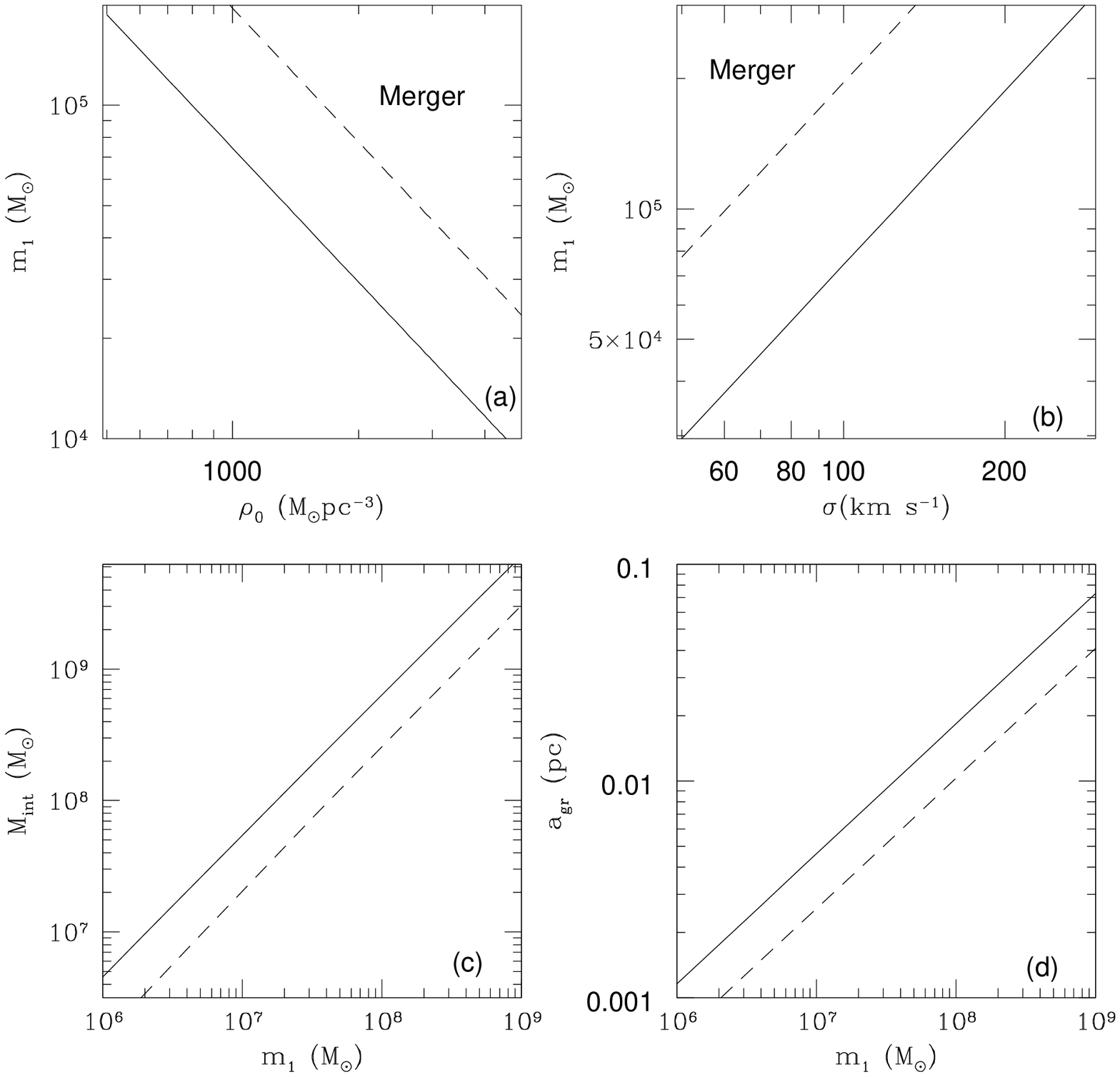}}
\caption{ (a) The line indicates those values of $m_1$ and galaxy
central density, $\rho_0$, for which the BBH reaches the gravitational
wave stage ($a=a_{gr}$) in 10 Gyr; points above this line correspond
to combinations for which this stage is reached in less time. The
stellar velocity dispersion $\sigma$ is set to 100 km/s.  (b) The line
indicates those values of $m_1$ and $\sigma$ for which the BBH reaches
$a=a_{gr}$ in 10 Gyr; points above this line correspond to
combinations for which this stage is reached in less time. $\rho_0=
1000 M_{\odot}\textrm{pc}^{-3}$.  (c) The total mass in stars the
binary must interact with until $a=a_{gr}$; $\sigma=$ 100 km/s and
$\rho_0=$~1000 $M_{\odot}\textrm{pc}^{-3}$.  (d) $a_{gr}$ as a
function of $m_1$; $\sigma=$~100 km/s and $\rho_0=$~1000
$M_{\odot}\textrm{pc}^{-3}$.  The solid lines are for $m_1=m_2$, the
dashed lines for $m_1= 10 m_2$.  }
\label{fig:4}
\end{figure}


\begin{references}

\reference{}
Aarseth, S.J. 2002, astro-ph/0210116

\reference{}
Aarseth, S.J. 1994, Direct Methods for N-Body Simulations, in Galactic
Dynamics and N-Body Simulations, eds. G. Contopoulos, N.K. Spyrou \&
L. Vlahos (Springer-Verlag)

\reference{}
Armitage, P.J. \& Natarajan, P. 2002, ApJ 567, L9

\reference{}
Begelman, M.C., Blandford, R.D. \& Rees, M.J. 1980, Nature 287, 307

\reference{}
Binney, J., \& Tremaine, S. 1987, Galactic Dynamics (Princeton:
Princeton Univ. Press) 

\reference{}
Byun, Y. et al. 1996, AJ 111, 1889

\reference{}
Chatterjee, P., Hernquist, L. \& Loeb, A. 2002a, ApJ 572, 371

\reference{}
Chatterjee, P., Hernquist, L. \& Loeb, A. 2002b, Phys. Rev. Lett. 88,
121103

\reference{}
Faber, S.M. et al. 1997, AJ 114, 1771

\reference{}
Frank, J. \& Rees, M.J. 1976, MNRAS 176, 633

\reference{}
Gould, A., \& Rix, H.-W. 2000, ApJ, 532, L29

\reference{}
Heggie, D.C. \& Mathieu, R.M. 1986, in P. Hut \& S.L.W. McMillan
(eds.), \emph{The Use of Supercomputers in Stellar Dynamics},
pp. 233-235, New York

\reference{}
Hemsendorf, M., Sigurdsson, S. \& Spurzem, R. 2002, astro-ph/0103410

\reference{}
Hernquist, L. \& Barnes, J. 1990, ApJ 349, 562

\reference{}
Hernquist, L. \& Ostriker, J.P. 1992, ApJ 386, 375

\reference{}
Hills, J.G. 1983, AJ 88, 1269

\reference{}
Hills, J.G. \& Fullerton, L.W. 1980, AJ 85, 1281

\reference{}
Lauer, T.R. et al. 1995, AJ 110, 2622

\reference{}
Magorrian, J., et al. 1998, AJ 115, 2285

\reference{}
Makino, J. 1997, ApJ 478, 58

\reference{}
Merritt, D. 2001, ApJ 556, 245

\reference{}
Merritt, D. \& Ferrarese, L. 2001, MNRAS 320, L30

\reference{}
Nakano, T. \& Makino, J. 1999a, ApJ 525, L77

\reference{}
Nakano, T. \& Makino, J. 1999b, ApJ 510, 155

\reference{}
Milosavljevi\'c, M. \& Merritt, D. 2001, ApJ 563, 34

\reference{}
Milosavljevi\'c, M. \& Merritt, D. 2002, astro-ph/0212459

\reference{}
Quinlan, G.D. 1996, NewA 1, 35

\reference{}
Quinlan, G.D. \& Hernquist, L. 1997, NewA 2, 533

\reference{}
Peters, P.C. 1964, Phys. Rev. B 136, 1224

\reference{}
Roos, N. 1988, ApJ 334, 95

\reference{}
Wandel, A. 1999, ApJ 519, L39

\reference{}
Yu, Q. 2002, MNRAS 331, 935

\end{references}
\end{document}